\documentstyle[nato,numreferences,epsfig]{crckapb} 

\newcommand{\Width}{6.2cm}
\newcommand{\qtext}[1]{\makebox[\Width][c]{\small\textbf{#1}}}
\newcommand{\qpic}[1]{\epsfig{file=q.#1.eps,width=\Width,%
bbllx=0,bblly=0,bburx=568,bbury=220}}

\begin{opening}
\title{DECORRELATION OF THE TOPOLOGICAL CHARGE IN \protect\\ 
TEMPERED SIMULATIONS OF FULL QCD}
\subtitle{}
\author{H.~St\"uben}
\institute{Konrad-Zuse-Zentrum f\"ur Informationstechnik Berlin \\
Takustrasse 7, 14195 Berlin, Germany}
\end{opening}

\begin{document}

\begin{abstract}
The improvement of simulations of QCD with dynamical Wilson fermions
by combining the Hybrid Monte Carlo algorithm with parallel tempering
is studied.  As an indicator for decorrelation the topological charge
is used.
\end{abstract}

\section{Introduction}

Decorrelation of the topological charge in Hybrid Monte Carlo (HMC)
simulations of QCD with dynamical fermions is a long standing problem.
For staggered fermions an insufficient tunneling rate of the
topological charge $Q_t$ has been observed \cite{MMP,Pisa}.  For Wilson
fermions the tunneling rate is adequate in many cases \cite{SESAM,RB1}.  
However on large lattices and for large values of $\kappa$ near the 
chiral limit the distribution of $Q_t$ is not symmetric even after more 
than 3000 trajectories (see Figure~1 of \cite{SESAM} and similar observations
by CP-PACS \cite{RB2}). 

It has also been observed that sensitive observables like the $\eta'$
correlator are $Q_t$ dependent \cite{KS}. Thus it appears to be important to
look for simulation methods that give good distributions of $Q_t$.

The idea of parallel tempering is to improve transitions in parameter
regions where tunneling is suppressed by opening ways through parameter
regions with little suppression. In QCD the method has
been applied successfully for staggered fermions \cite{Boyd}. In
\cite{UKQCD} parallel tempering has been used to simulate QCD with
O($a$)-improved Wilson fermions without finding any gain, however, 
with only two ensembles which does not take advantage of the main idea 
of the method.

Here parallel tempering is used in conjunction with HMC to simulate
QCD with (standard) Wilson fermions.  The gain achieved is
demonstrated by studying time series and histograms of the topological
charge and by comparing statistical errors of the topological
susceptibility $\langle Q_t^2\rangle$.

\section{Parallel Tempering}

In standard Monte Carlo simulations one deals with one parameter set 
$\lambda$ and generates a sequence of configurations $C$. The set $\lambda$ 
here includes $\beta$, $\kappa$, the leapfrog time step and the number of 
time steps. $C$ comprises the gauge field and the pseudo fermion field.

In the parallel tempering approach \cite{HN,EM} one simulates $N$ 
ensembles ($\lambda_i; C_i),\,i = 1, \dots, N$ in a
single combined run.  Two steps alternate: (a) update of $N$
configurations in the standard way, (b) exchange of configurations by
swapping pairs. Swapping of a pair of configurations means
\begin{equation}
((\lambda_i; C_i), (\lambda_j; C_j)) \rightarrow \left\{
\begin{array}{ll}
\!((\lambda_i; C_j), (\lambda_j; C_i)), & \!\mbox{if accepted} \\
\!((\lambda_i; C_i), (\lambda_j; C_j)), & \!\mbox{else}
\end{array}
\right.
\end{equation}
with the Metropolis acceptance condition 
\begin{equation}
P_{\rm swap}(i,j) = \min\left( 1, e^{-\Delta H} \right) \,,
\label{Pswap}
\end{equation}
\begin{equation}
\Delta H = 
H_{\lambda_i}(C_i) + H_{\lambda_j}(C_j) -
H_{\lambda_i}(C_j) - H_{\lambda_j}(C_i).
\end{equation}
Since after swapping both ensembles remain in equilibrium, the
swapping sequence can be freely chosen.  In order to achieve a high
swap acceptance rate one will only try to swap $(\beta,\kappa)$-pairs
that are close together.  If the chosen $(\beta,\kappa)$-values lie on
a curve in the $(\beta,\kappa)$-plane there are three obvious choices
for the swapping sequence of neighboring $(\beta,\kappa)$-pairs.  One
can step through the curve in either direction or swap randomly.  It
has turned out that it is advantageous to step along such a curve in
the direction from high to low tunneling rates of $Q_t$.

\section{Simulation Details}

The standard Wilson action for the gauge and the fermion fields was
used.  The lattice size was $8^4$.  The HMC program applied the
standard conjugate gradient inverter with even/odd
preconditioning. The trajectory length was always~1. The time steps
were adjusted to get acceptance rates of about 70\%. In all cases 1000
trajectories were generated (plus 50--100 trajectories for
thermalization).

$Q_t$ was measured by the field-theoretic method after 50 cooling steps of
Cabibbo-Marinari type. This method gives close to integer values which
were rounded to the nearest integers. (Note that the results presented
in \cite{lat99} were obtained without rounding.)

Statistical errors were obtained by binning, i.e., the values given are
the maximal errors calculated after blocking the data into bins of
sizes $10,20,50$ and $100$.

\section{Results}

Several tempered HMC simulations were run in the quenched approximation
(tempering in $\beta$) and with dynamical fermions (tempering in
$\kappa$, at fixed $\beta=5.5$ and $\beta=5.6$).  For comparison also
standard HMC simulations have been performed. 

\begin{figure}[h]
\qtext{Standard HMC}\hspace{0.1cm}\qtext{Tempered HMC}

\medskip

\qpic{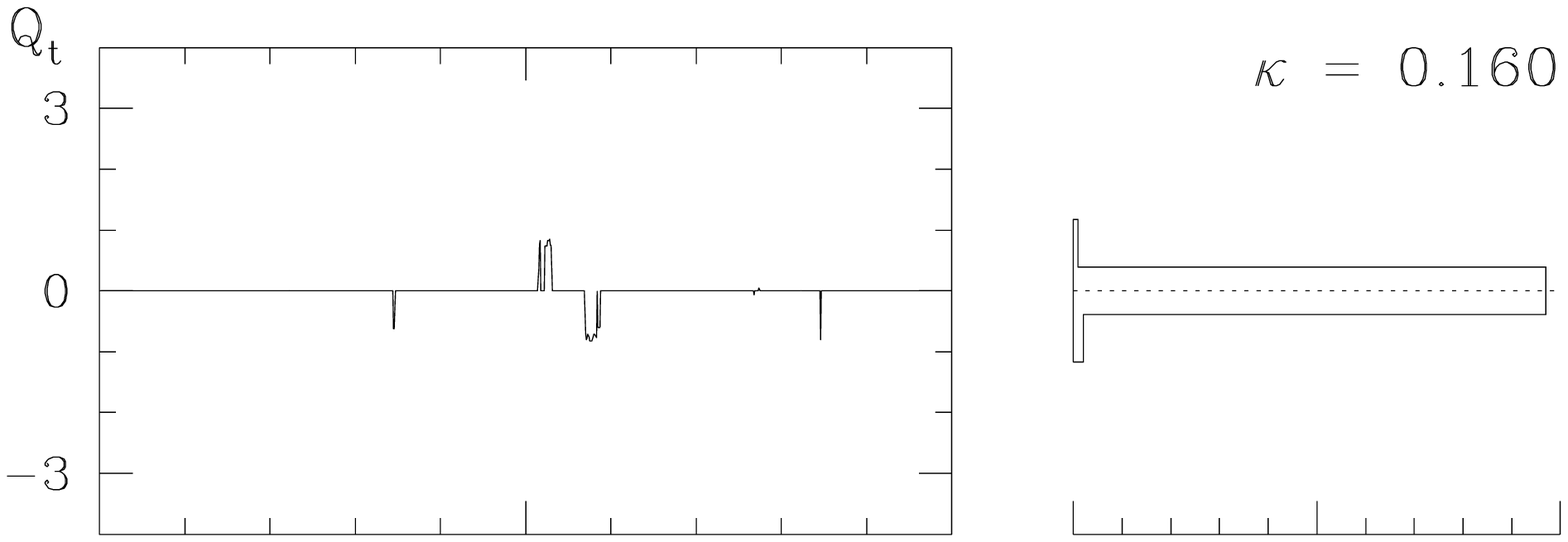}\hspace{0.1cm}\qpic{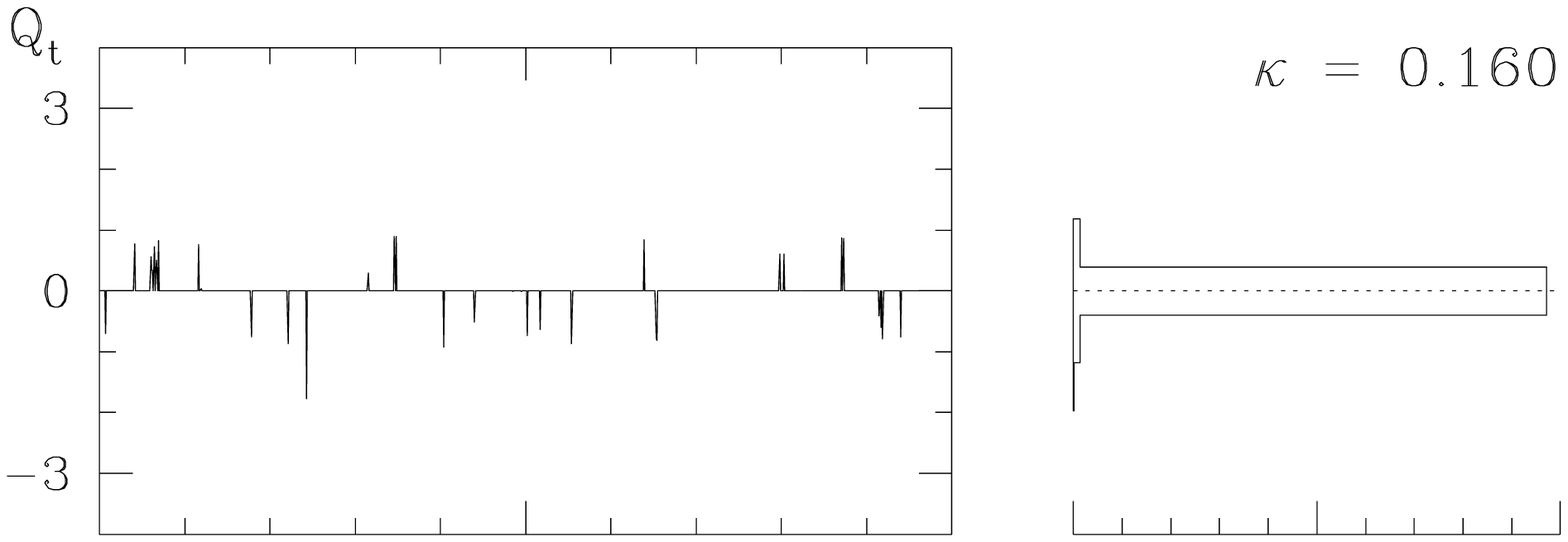}
\qpic{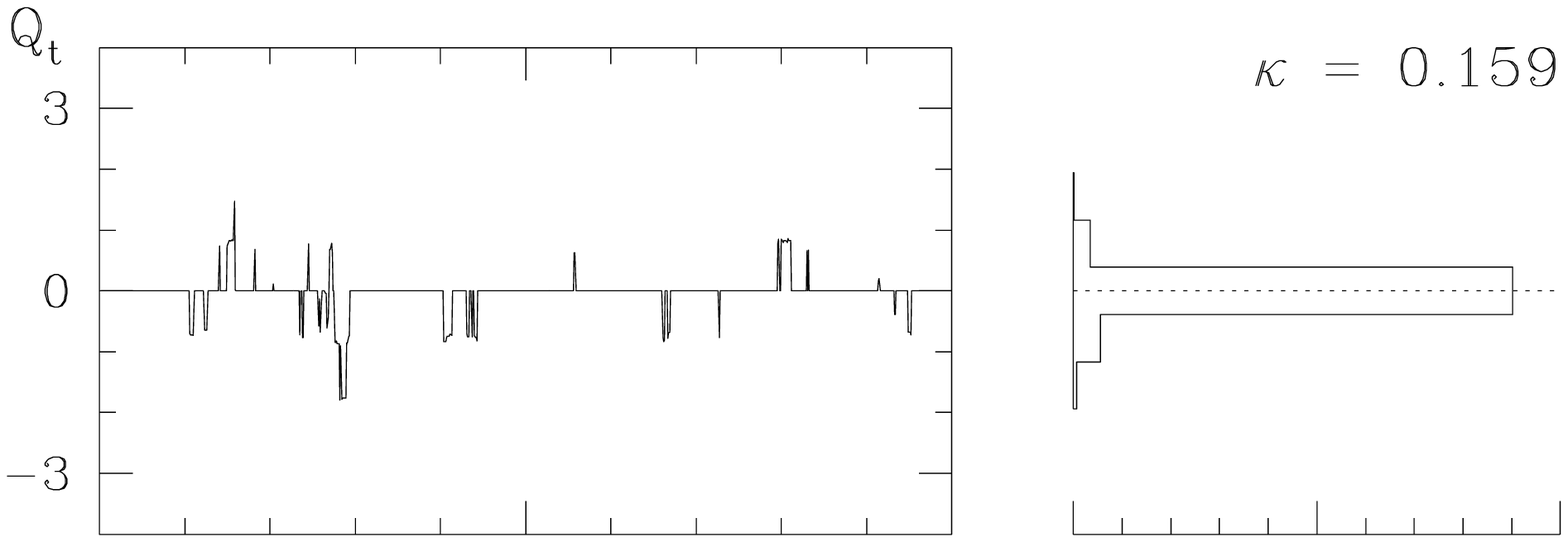}\hspace{0.1cm}\qpic{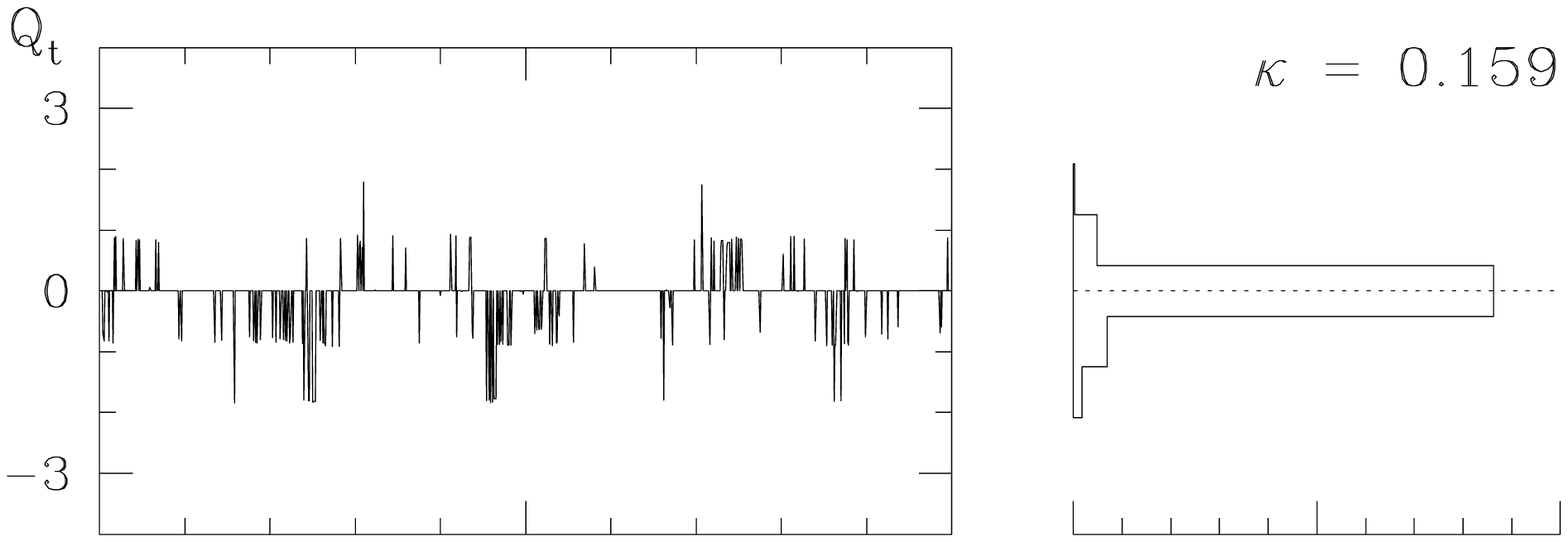}
\qpic{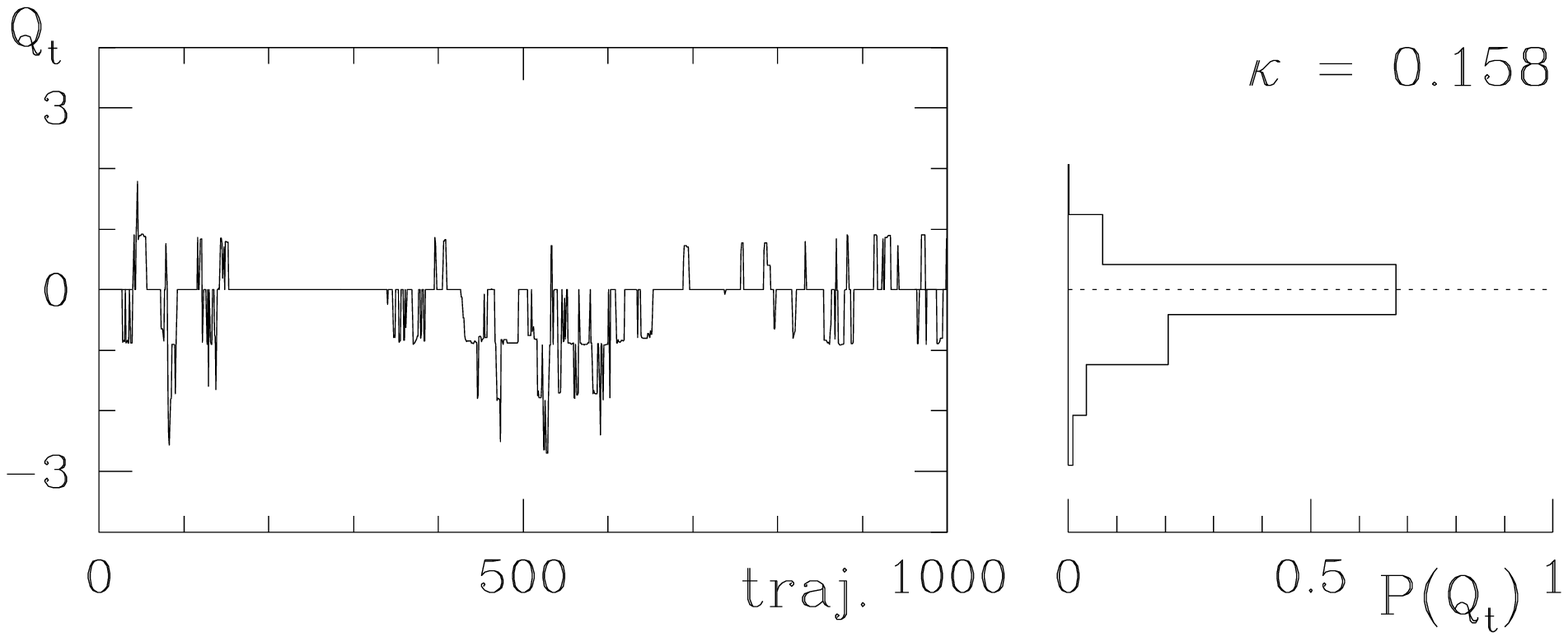}\hspace{0.1cm}\qpic{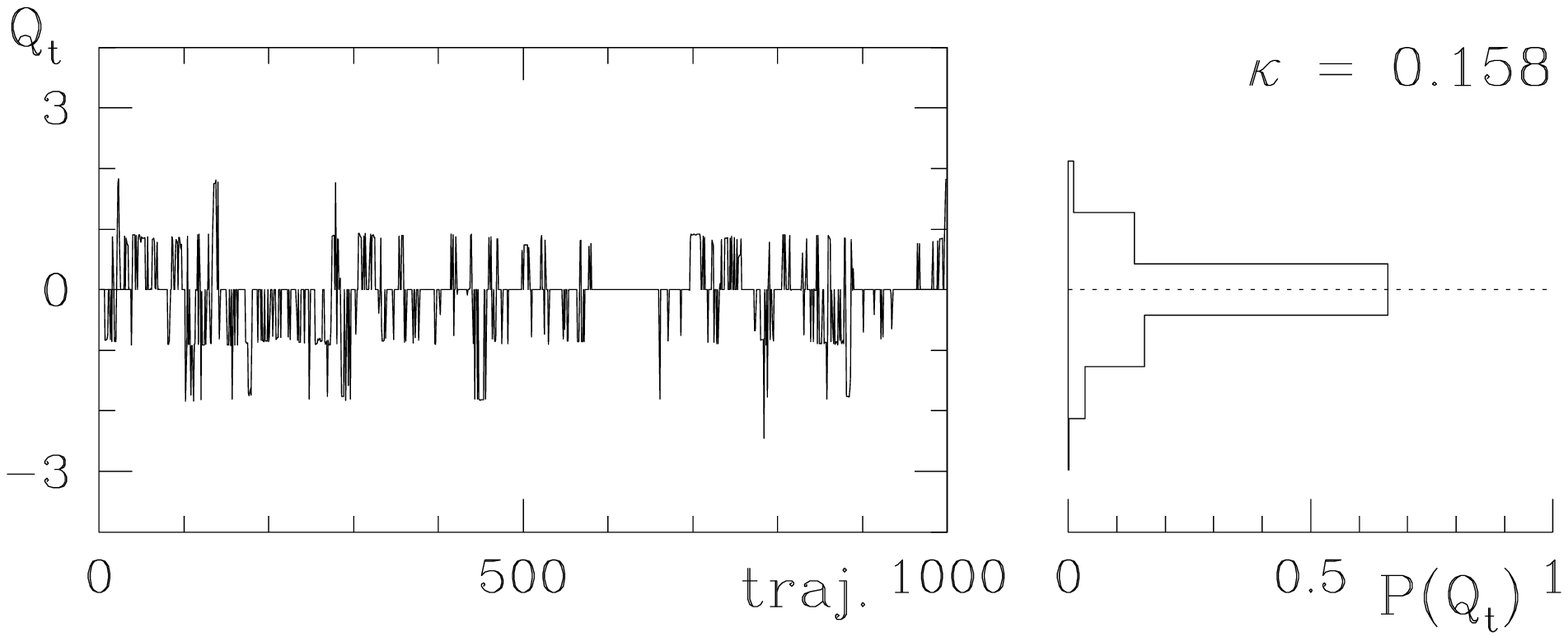}

\caption{\label{fig5.5}Comparison of time series and histograms of
$Q_t$ obtained from standard and tempered HMC on the $8^4$ lattice at
$\beta = 5.5$. In the tempered run 5 ensembles were used, $0.158 \leq
\kappa \leq 0.160$ and $\Delta\kappa = 0.0005$. The swap acceptance
rate was about 56\%.}
\end{figure}

\begin{figure}[t]
\qtext{Standard HMC}\hspace{0.1cm}\qtext{Tempered HMC}

\medskip

\qpic{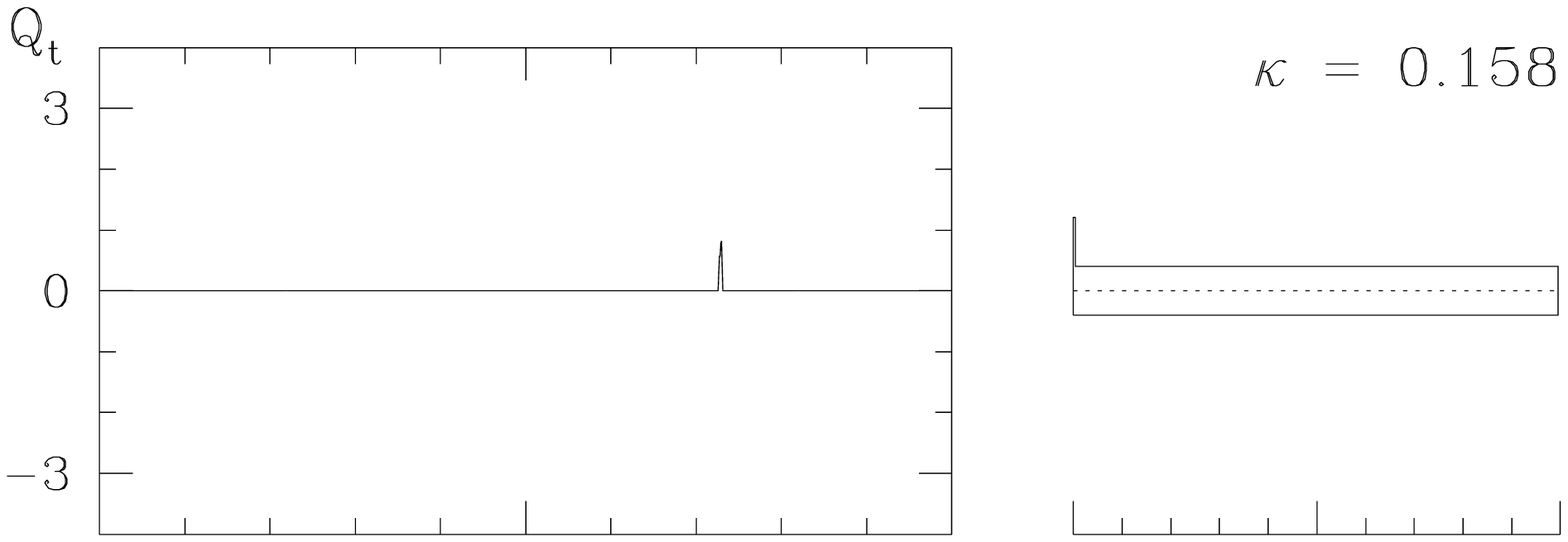}\hspace{0.1cm}\qpic{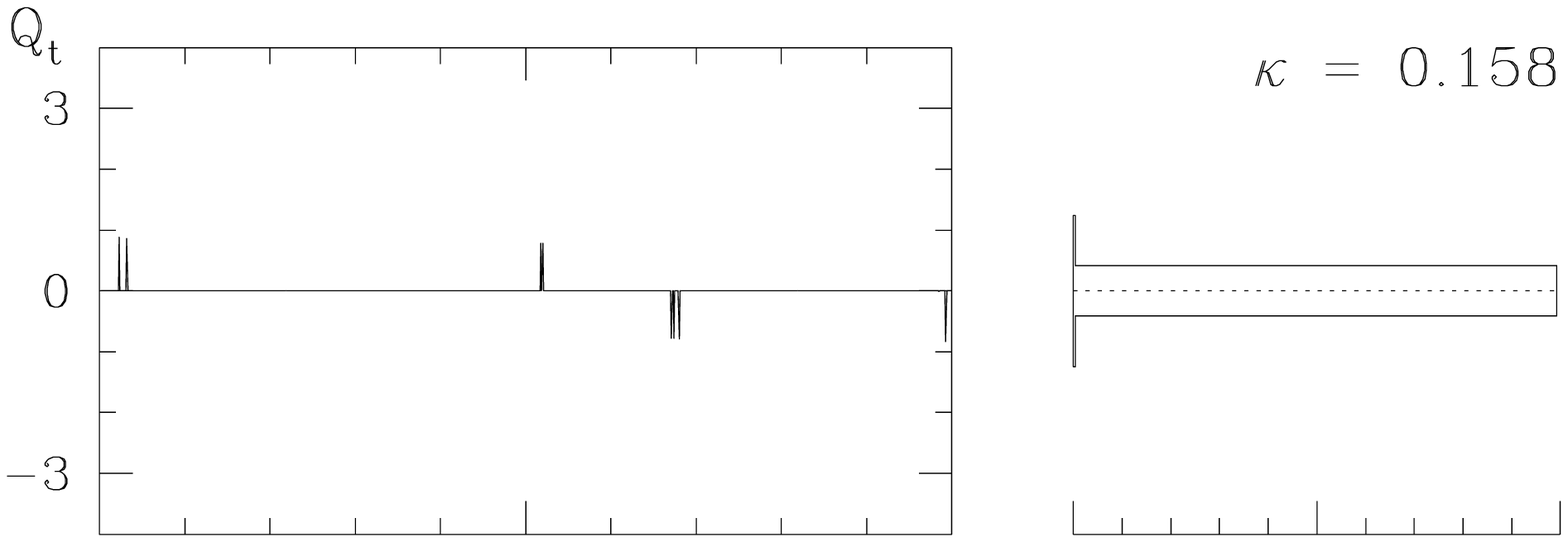}
\qpic{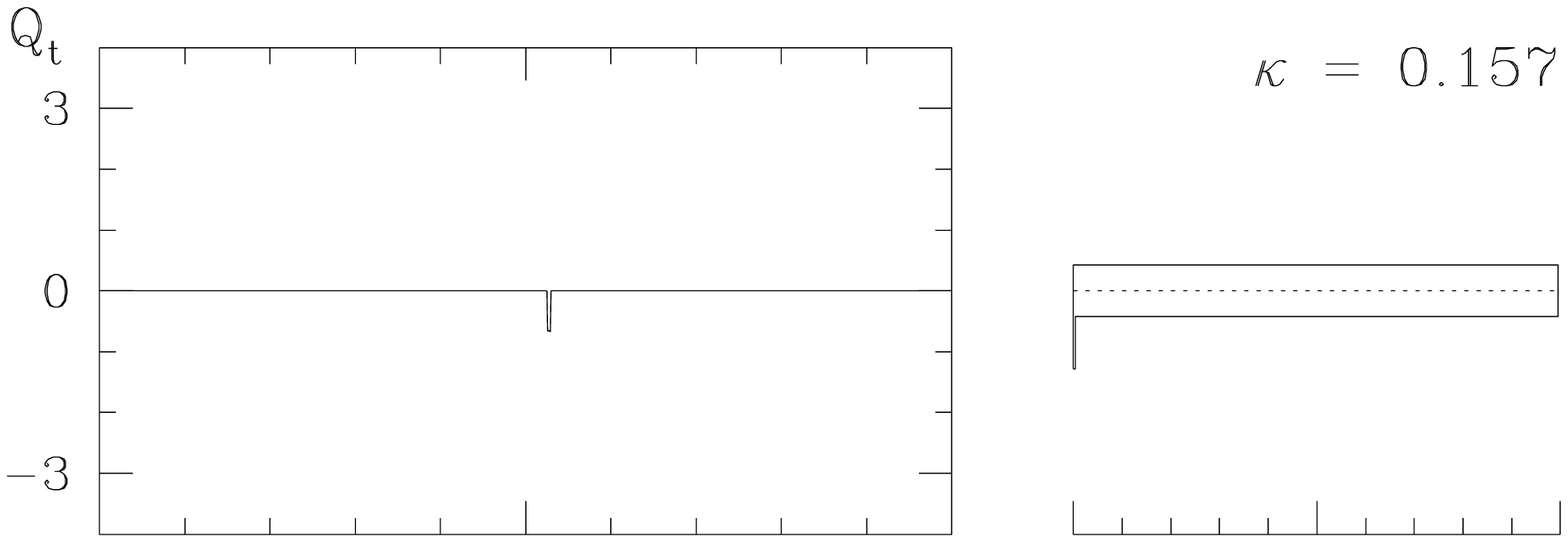}\hspace{0.1cm}\qpic{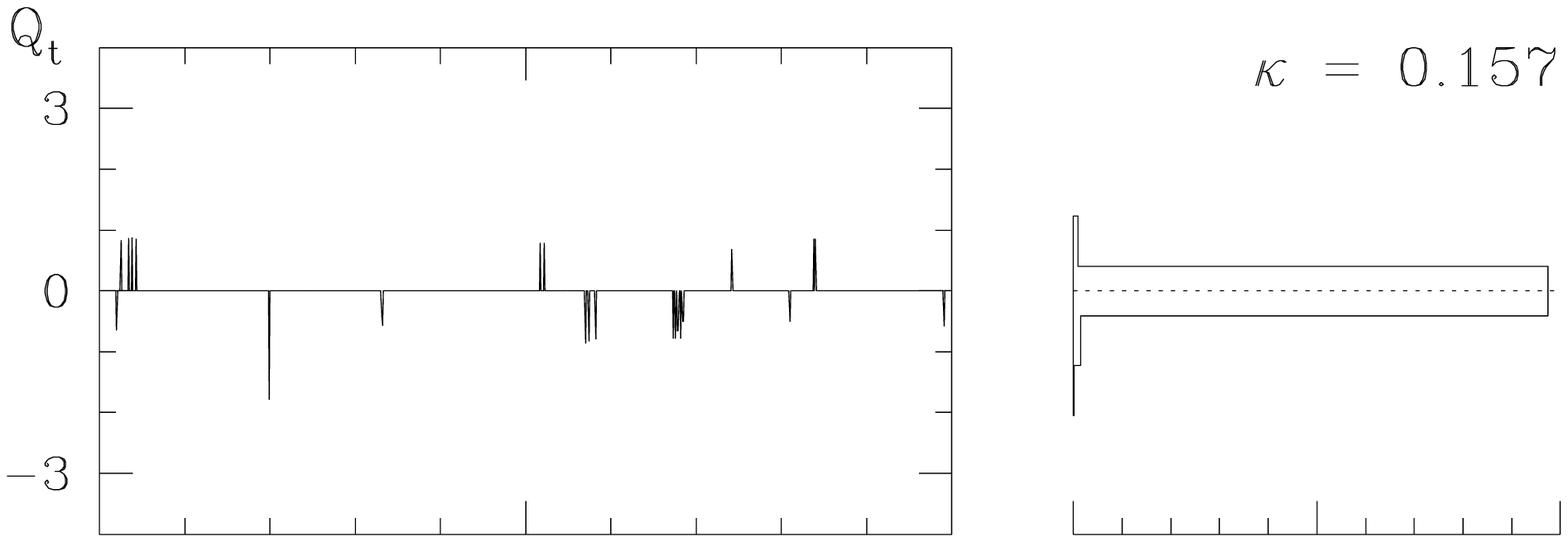}
\qpic{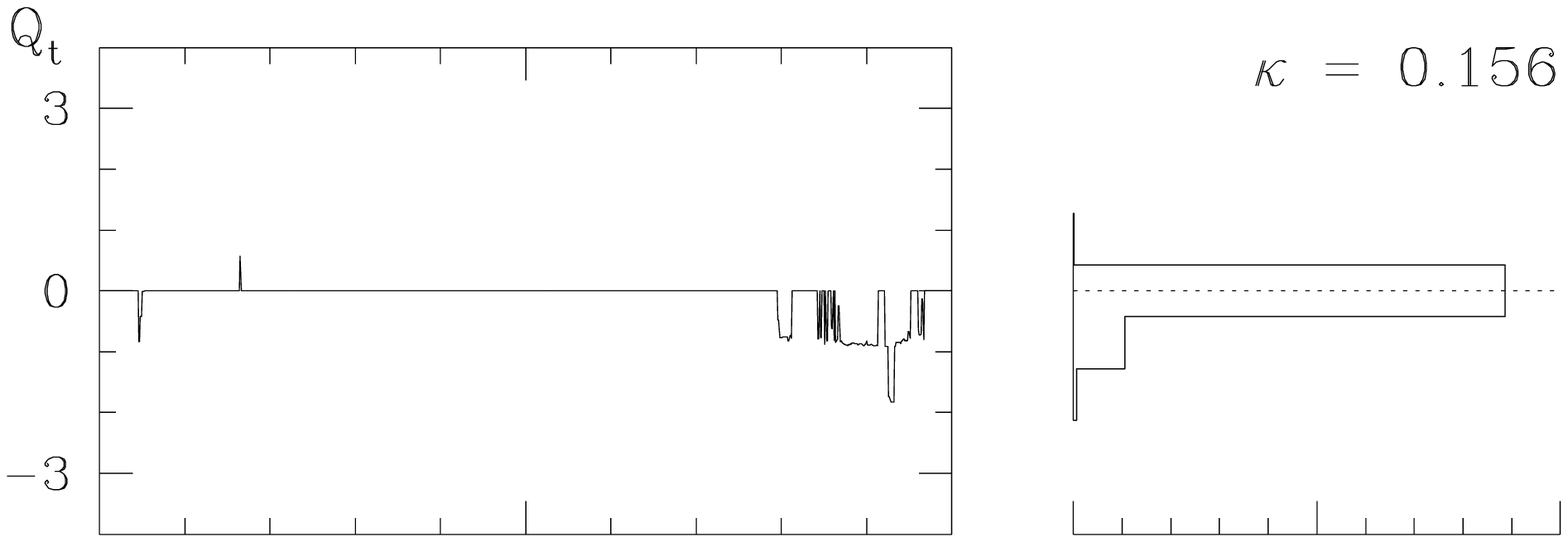}\hspace{0.1cm}\qpic{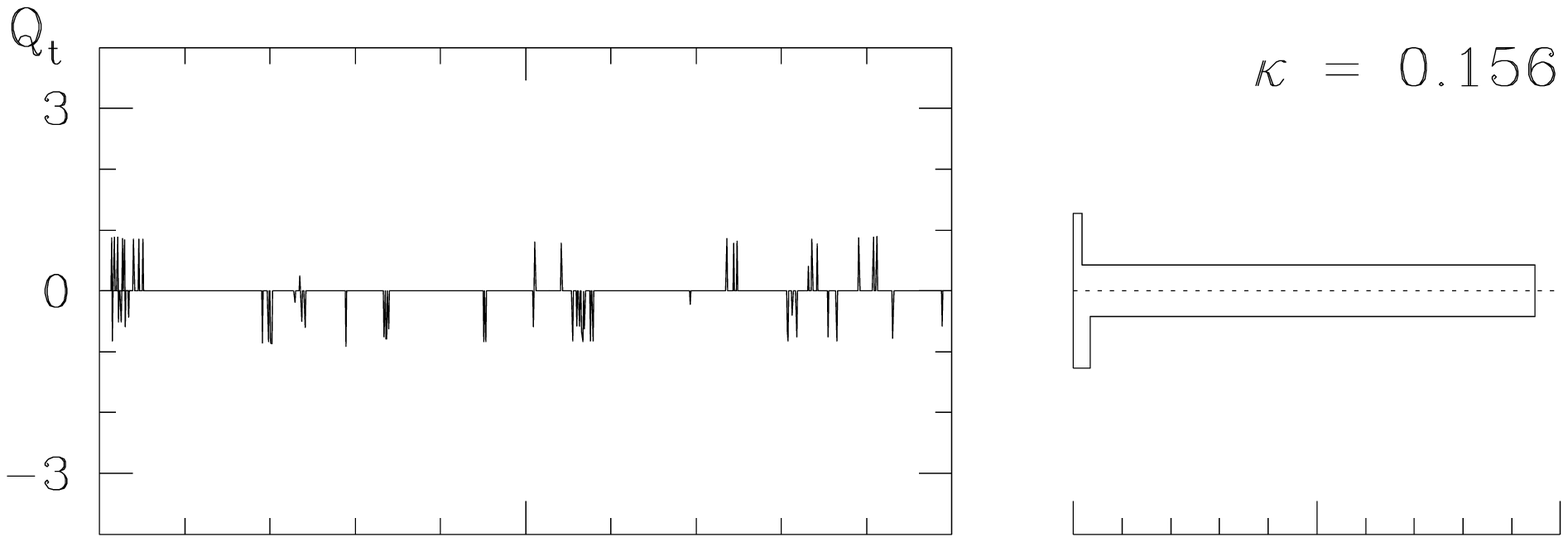}
\qpic{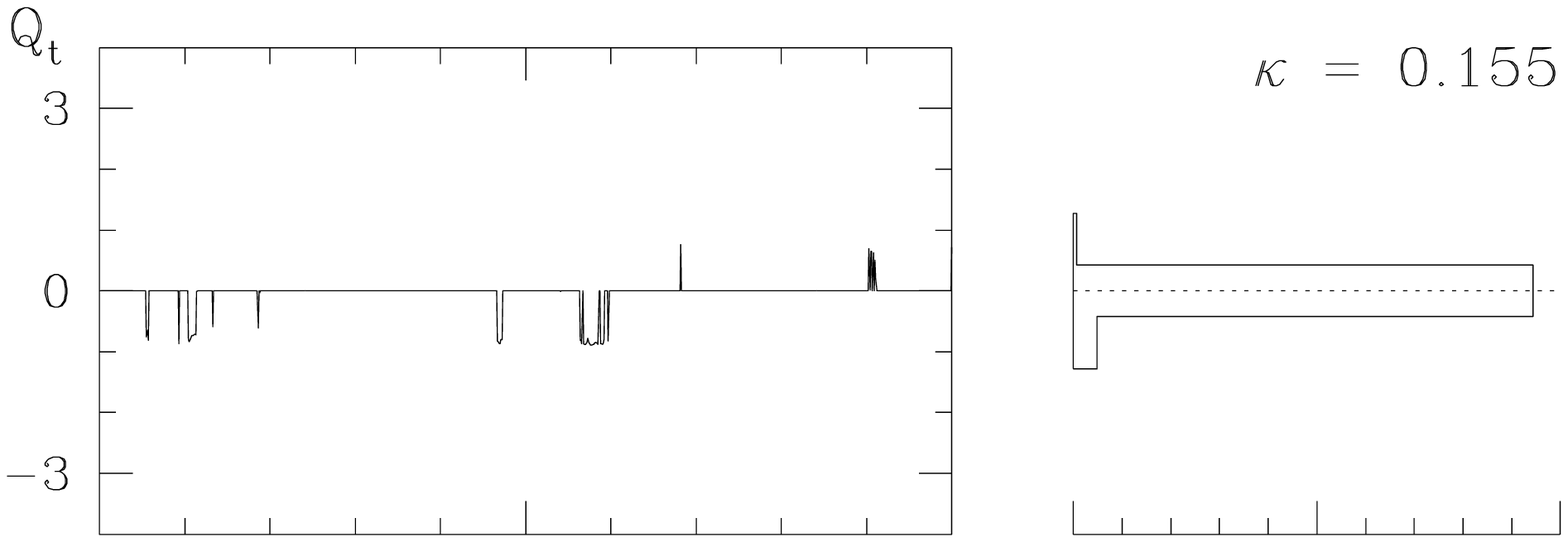}\hspace{0.1cm}\qpic{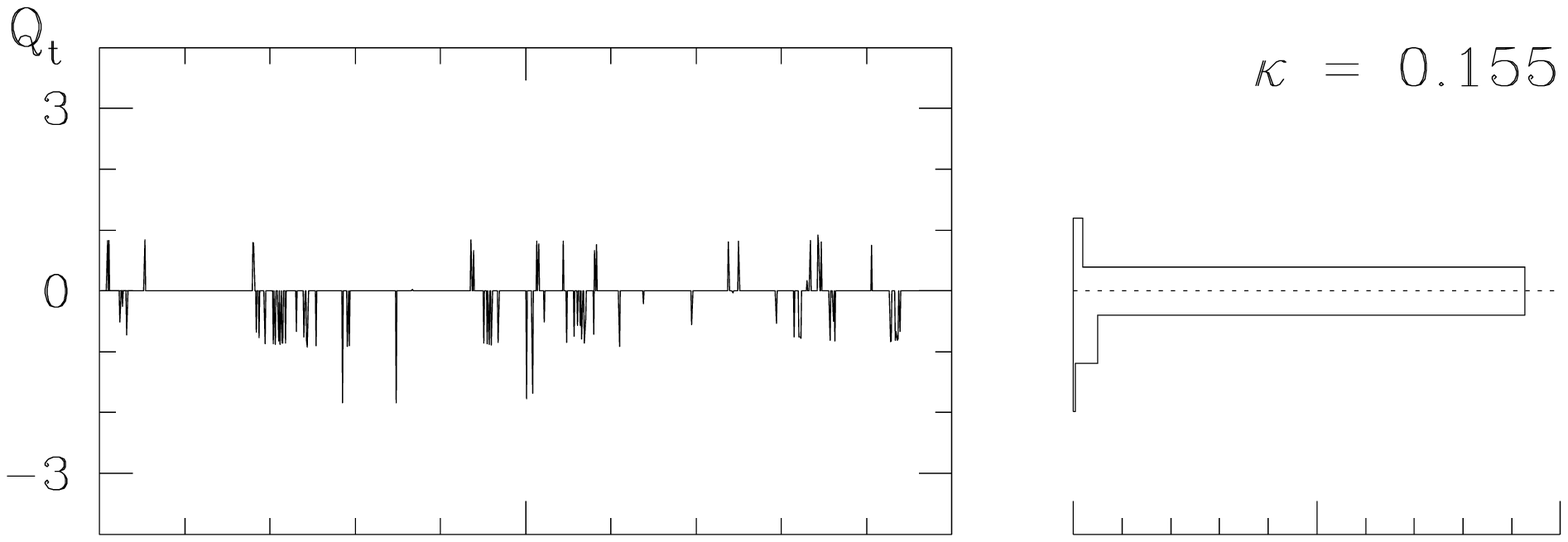}
\qpic{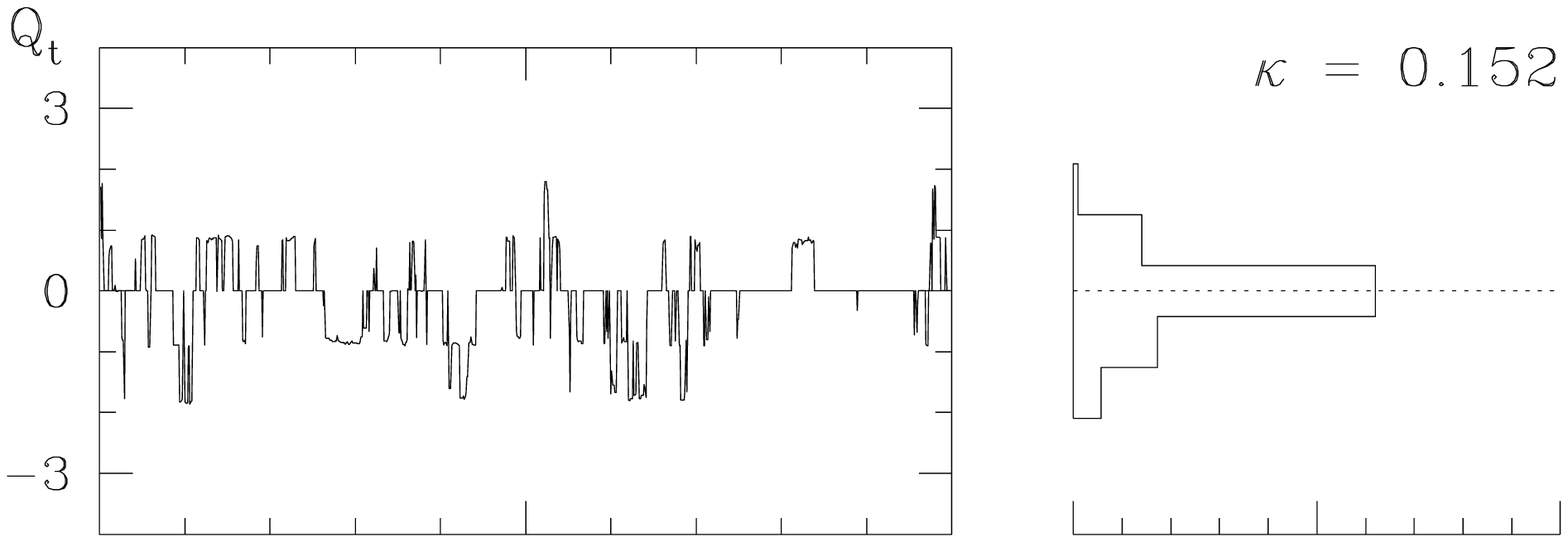}\hspace{0.1cm}\qpic{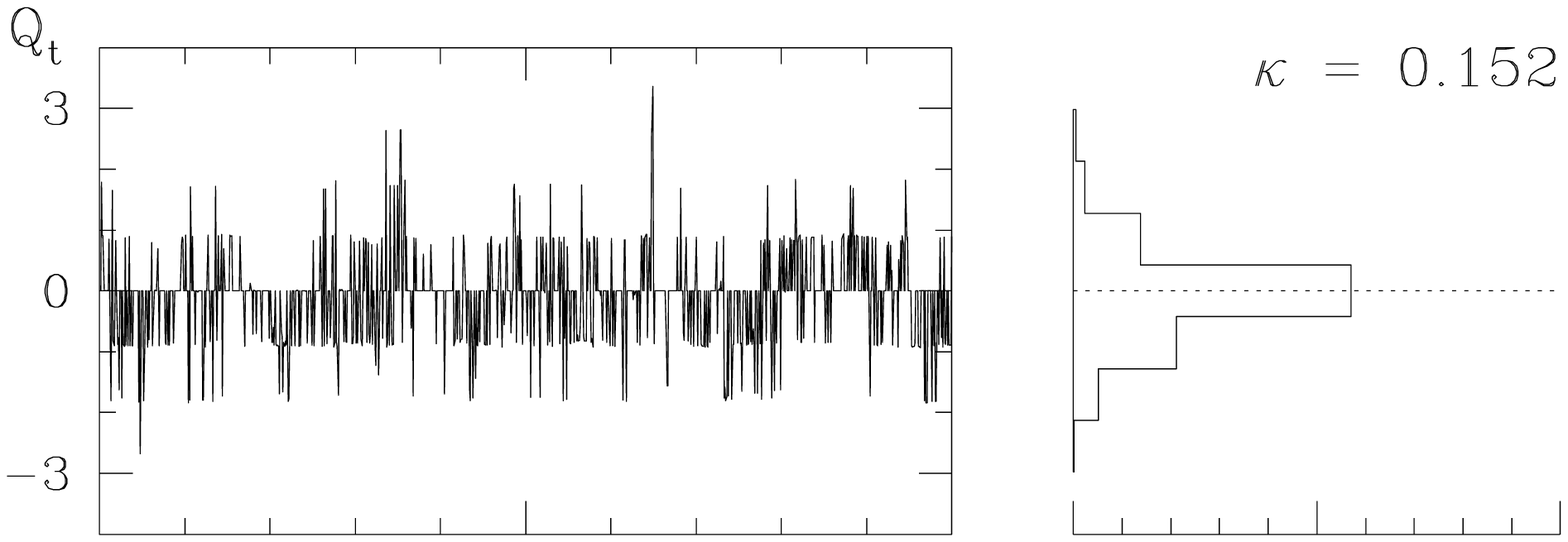}
\qpic{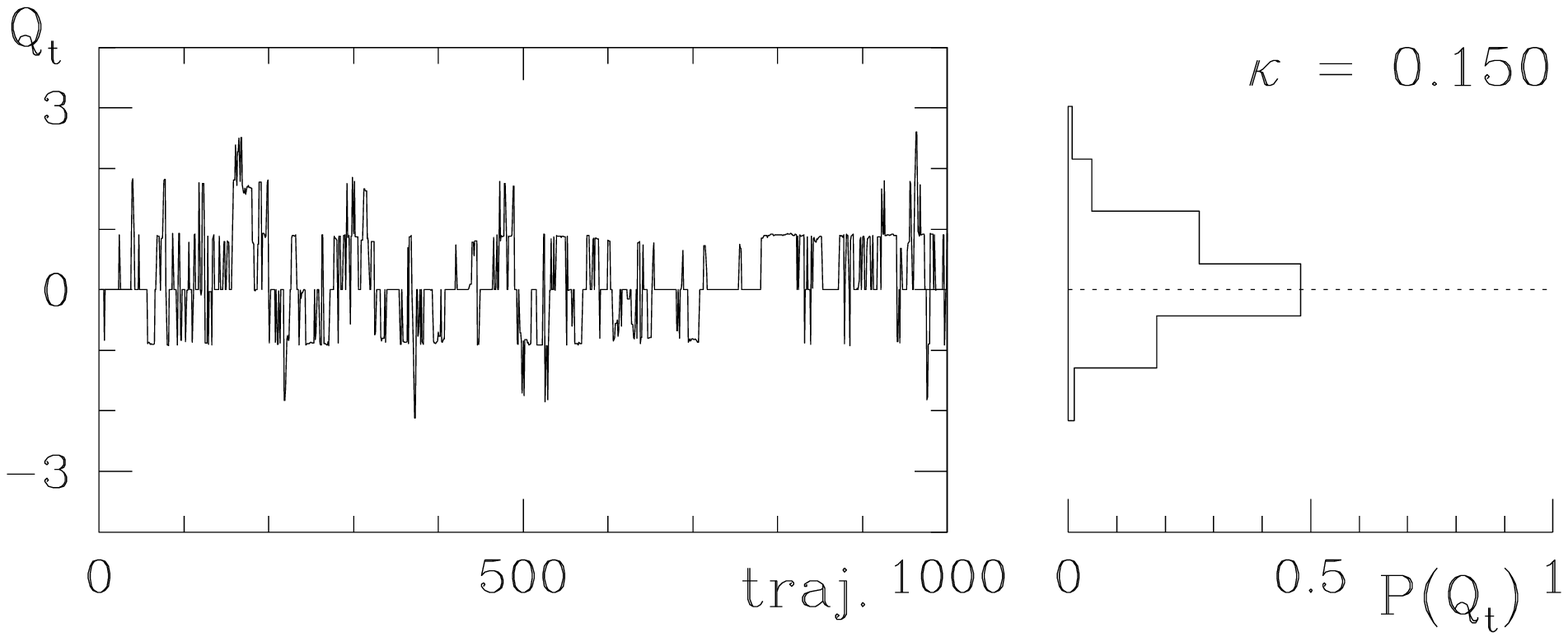}\hspace{0.1cm}\qpic{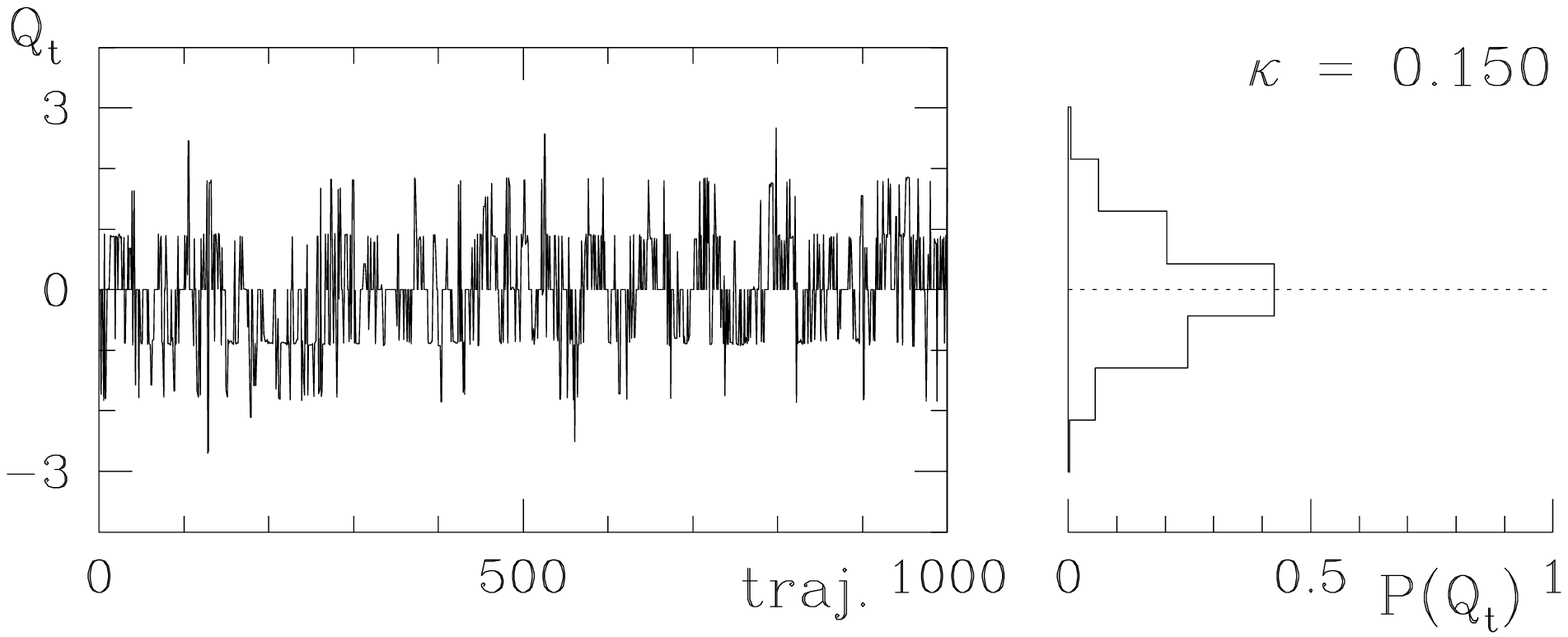}

\caption{\label{fig5.6}Comparison of time series and histograms of
$Q_t$ obtained from standard and tempered HMC on the $8^4$ lattice at
$\beta=5.6$.  The corresponding quantitative results can be found in
Table~\ref{tab5.6}.}
\end{figure}

Figures \ref{fig5.5} and \ref{fig5.6} show typical comparisons of time
series and histograms of $Q_t$.  One sees that with tempering
considerably more topologically nontrivial configurations occur and
that the histograms of $Q_t$ become in general more symmetrical and
broader.

In standard runs $Q_t$ frequently stayed for quite some time near 1 or
near $-1$, while with tempering this never occurred. The standard run
at $\kappa = 0.156$ shown in Figure~\ref{fig5.6}, where $Q_t$ gets
trapped in this way for about 200 trajectories, provides an example of
this. Such observations have also been made on large lattices
\cite{SESAM,RB2}.

While a correlation analysis cannot be carried out with the given size of
samples, some quantitative account of the improvement by tempering is
possible using the mean of the absolute change of $Q_t$,
called mobility in \cite{SESAM},
\begin{equation}
D_1 = \frac{1}{N_{\rm traj}} \sum_{i=1}^{N_{\rm traj}}
\left| Q_t(i) - Q_t(i - 1) \right |\;.
\label{D1}
\end{equation}
Results for $D_1$ are given in Tables~\ref{tab5.5} and~\ref{tab5.6}.
If $|Q_t(i) - Q_t(i - 1)| \leq 1$ for all trajectories then $1/D_1$ is
the HMC time between topological events.  Since that condition holds
in most of the cases presented here one gets an idea of the
quantitative improvement by tempering.

Another quantitative estimate of improvement comes from the
statistical errors of $\langle Q_t^2\rangle$. The fact that
statistical errors decrease with the square of HMC time provides a
second quantitative criterion for the speed-up of a simulation.

Quantitative results at $\beta = 5.5$ are summarized in
Table~\ref{tab5.5}.  From the ratios of mobilities and squared ratios
of errors of susceptibilities one obtains speed-ups between 2 (ratio
of $D_1$ at $\kappa = 0.158$) and 16 (squared ratio of the errors of
$\langle Q_t^2\rangle$ at $\kappa = 0.160$).  This is a considerable
gain, especially if runs at several values of $\kappa$ need to be
done, what is usually the case.

\begin{figure}[h]
\begin{center}
\epsfig{file=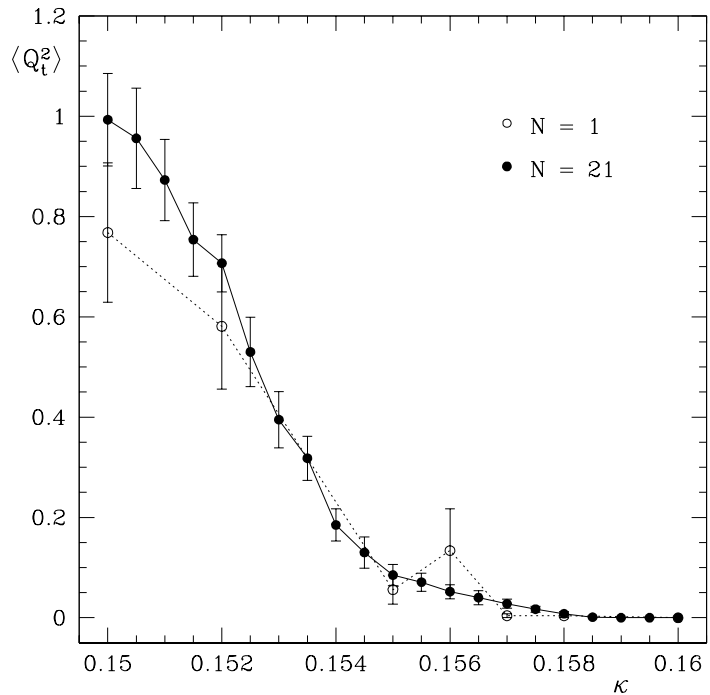,width=10cm}
\vspace{-6mm}
\end{center}
\caption{\label{fig5.6_Q2}Comparison of topological susceptibilities
on the $8^4$ lattice at $\beta = 5.6$. The plot shows results from
standard HMC ($N = 1$) and the tempered HMC run with $N = 21$
ensembles.}
\end{figure}

At $\beta = 5.6$ tempering looks even better in the sense that the
standard HMC runs do not really resolve the topological properties for
$\kappa \geq 0.156$ (see Table~\ref{tab5.6} and Figures~\ref{fig5.6}
and~\ref{fig5.6_Q2}).
\pagebreak

\vspace*{-6mm}

\begin{table}[h]
\begin{center}
\caption{\label{tab5.5}Mobilities $D_1$ and topological
susceptibilities $\langle Q_t^2\rangle$ for the plots shown in
Figure~\ref{fig5.5}.}
\begin{tabular}{lllll}
\hline
& \multicolumn{2}{c}{Standard HMC} 
& \multicolumn{2}{c}{Tempered HMC} \\
\multicolumn{1}{c}{$\kappa$} & 
\multicolumn{1}{c}{$D_1$} & 
\multicolumn{1}{c}{$\langle Q_t^2\rangle$} & 
\multicolumn{1}{c}{$D_1$} & 
\multicolumn{1}{c}{$\langle Q_t^2\rangle$} \\
\hline
0.158 & 0.171(35) & 0.51(19)  & 0.398(53) & 0.49(8) \\
0.159 & 0.058(20) & 0.12(5)   & 0.248(40) & 0.20(5) \\
0.160 & 0.012(8)  & 0.030(27) & 0.056(13) & 0.031(7) \\
\hline
\end{tabular}
\end{center}
\end{table}

\vspace*{-2mm}

In the following the choice of $\kappa$-values at $\beta = 5.6$ is
motivated. The run with 21 ensembles can be considered as a reference
run. In a large scale simulation one would want to use less
ensembles. The run with 6 ensembles demonstrates that comparable
speed-up can be achieved with a smaller number of ensembles. The
run with 7 ensembles covers exactly the parameter range investigated
by SESAM \cite{SESAM}. It was mainly done to get estimates for the swap
acceptance rate on larger lattices for $\Delta\kappa = 0.00025$ (see
section~\ref{secBig}).

It is interesting to compare the runs with 6 and 7 ensembles. In the
run with 6 ensembles the mobility is higher. This reflects the main
idea of the tempering method which is to connect areas of low
tunneling rates with areas of high tunneling rates. 
\clearpage

\clearpage

\begin{table}[h]

\caption{\label{tab5.6}Mobilities $D_1$ and topological
susceptibilities $\langle Q_t^2\rangle$ on the $8^4$ lattice at
$\beta=5.6$. The swap acceptance rates achieved were
about 82\% for $\Delta\kappa = 0.00025$ and about 63\% for
$\Delta\kappa = 0.0005$.}

\begin{tabular}{lllll}
\hline
  & Standard HMC & \multicolumn{3}{c}{Tempered HMC} \\
  &              & 7 ensembles  & 6 ensembles & 21 ensembles \\
  &              & $0.156 \leq \kappa \leq 0.1575$ 
                 & $0.155 \leq \kappa \leq 0.1575$
                 & $0.15 \leq \kappa \leq 0.16$ \\
  &              & $\Delta\kappa = 0.00025$ 
                 & $\Delta\kappa = 0.0005$ 
                 & $\Delta\kappa = 0.0005$ \\
\hline
\multicolumn{1}{c}{$\kappa$} & 
\multicolumn{4}{c}{$D_1$} \\
\hline
0.1500 & 0.325(39)&          &           & 0.764(46)\\
0.1520 & 0.174(29)&          &           & 0.735(50)\\[0.7ex]
0.1550 & 0.031(11)&          &  0.167(42)& 0.132(32)\\
0.1555 &          &          &  0.176(43)& 0.118(30)\\
0.1560 & 0.030(17)& 0.064(27)&  0.108(36)& 0.096(26)\\     
0.1565 &          & 0.040(15)&  0.102(28)& 0.074(22)\\
0.1570 & 0.002(2) & 0.022(8) &  0.068(19)& 0.046(15)\\
0.1575 &          & 0.004(3) &  0.044(14)& 0.034(12)\\
0.1580 & 0.002(2) &          &           & 0.016(8)\\[0.7ex]
0.1600 & 0        &          &           & 0 \\
\hline
\multicolumn{1}{c}{$\kappa$} & 
\multicolumn{4}{c}{$\langle Q_t^2\rangle$} \\
\hline
0.1500 & 0.77(14) &          &          & 0.993(92)\\
0.1520 & 0.58(13) &          &          & 0.707(57)\\[0.7ex]
0.1550 & 0.056(29)&          & 0.144(40)& 0.085(21)\\
0.1555 &          &          & 0.100(25)& 0.071(18)\\
0.1560 & 0.134(83)& 0.044(20)& 0.062(23)& 0.052(14)\\     
0.1565 &          & 0.020(8) & 0.055(17)& 0.040(14)\\
0.1570 & 0.004(4) & 0.011(4) & 0.037(11)& 0.028(9)\\
0.1575 &          & 0.002(1) & 0.030(11)& 0.017(6)\\
0.1580 & 0.004(4) &          &          & 0.008(4)\\[0.7ex]
0.1600 & 0        &          &          & 0 \\
\hline
\end{tabular}
\end{table}

\vspace*{-5mm}

\section{Going to larger Lattices} \label{secBig}

With regard to large scale simulations of QCD performance predictions
are needed.  One potential problem of the tempering method has been
stressed in \cite{UKQCD}, namely the decrease of the swap acceptance
rate $\langle A\rangle$ with the lattice volume.  In \cite{UKQCD} it
has been checked that the relation \cite{Bielefeld}
\begin{equation} \label{eqA}
\langle A \rangle = \mbox{erfc} \left( \frac{1}{2} 
\sqrt{\langle \Delta H \rangle} \right)
\end{equation}
is valid for a large range of $\langle \Delta H \rangle$. Relation
(\ref{eqA}) also holds in all simulation done in this work.

\pagebreak

Because $\langle \Delta H \rangle$ scales linearly with the lattice
volume $V$, relation (\ref{eqA}) allows one to predict $\langle A
\rangle$ by inserting values measured on the $8^4$ lattice.
Table~\ref{tabA} lists predictions using values of $\langle A \rangle$
from the runs shown in Table~\ref{tab5.6}. Some caution is necessary
with these predictions because on the $8^4$ lattice at $\beta = 5.6$
and $0.15 \leq \kappa \leq 0.16$ the finite temperature phase
transition \cite{PT} is crossed.

\vspace{-8mm}

\begin{table}[h]
\begin{center}
\caption{\label{tabA}}
\begin{tabular}{lll}
\hline
\multicolumn{1}{c}{$\Delta\kappa$} & 
\multicolumn{1}{c}{$V$} & 
\multicolumn{1}{c}{$\langle A\rangle$} \\
\hline
0.0005  & $8^4$            & 63\% \\
        & $16^3 \times 32$ & 0.6\% \\[0.7ex]
0.00025 & $8^4$            & 82\% \\
        & $16^3 \times 32$ & 20\% \\
        & $24^3 \times 48$ & 0.4\% \\

\hline
\end{tabular}
\end{center}
\end{table}

\vspace{-4mm}

Indeed more and more ensembles will be needed on larger lattices if
one wants to keep $\langle A\rangle$ and the parameter range constant.
However it is an open
question which effect is stronger, the decrease
of $\langle A \rangle$ or the slowing down of tunneling between
topological sectors. The hope is that the need to take more ensembles
more than compensates the slowing down of tunneling.

\section{Conclusions}

On the $8^4$ lattice parallel tempering considerably enhances
tunneling between different sectors of topological charge and
generates samples with more symmetrical charge distributions than can
be obtained by standard HMC. The histograms also get slightly broader
or even become nontrivial thanks to this technique.

The enhancement of tunneling indicates an improvement of decorrelation
also for other observables. More satisfactory histograms are important
for topologically sensitive quantities. Both of these features make
parallel tempering an attractive method for large-scale QCD
simulations. The method is particularly economical when several
parameter values have to be studied anyway.

A potential problem is that for a given parameter set the swap
acceptance rate (\ref{Pswap}) decreases for increasing lattice volume
\cite{UKQCD}.  To settle the question whether on larger lattices the
need for increasing the number of ensembles is compensated by improved
tunneling between topological sectors this study will be continued on
larger lattices.

\section*{Acknowledgements}

This work was done in collaboration with E.-M.~Ilgenfritz and
W.~Kerler.  I would like to thank M.~M\"uller-Preussker for
supporting the project. The simulations were done on the CRAY T3E at
Konrad-Zuse-Zentrum f\"ur Informationstechnik Berlin.

\end{document}